\documentclass[12pt]{scrartcl}

\usepackage{abstract}
\usepackage{graphicx}
\usepackage{wrapfig}
\usepackage{placeins}
\usepackage{hyperref}
\usepackage[round]{natbib}

\begin{document} 

\title{Context-Aware Prediction of Pathogenicity of Missense Mutations Involved in Human Disease}
\author{Christoph Feinauer\textsuperscript{1}, Martin Weigt\textsuperscript{1}}
\date{\small{1 Sorbonne Universit\'{e}s, UPMC, Institut de Biologie
  Paris-Seine, CNRS, Laboratoire de Biologie Computationnelle et
  Quantitative UMR 7238, Paris, France}}

\maketitle
\begin{onecolabstract}
Amino-acid substitutions are implicated in a wide range of human diseases, many
of which are lethal.  Distinguishing such mutations from polymorphisms without significant
effect on human health is a necessary step in understanding the etiology of
such diseases. Computational methods can be used to select interesting
mutations within a larger set, to corroborate experimental findings and to
elucidate the cause of the deleterious effect. In this work, we show that
taking into account the sequence context in which the mutation appears allows
to improve the predictive and explanatory power of such methods. We present an
unsupervised approach based on the direct-coupling analysis of homologous
proteins. We show its capability to quantify mutations where methods without
context dependence fail.  We highlight cases where the context dependence is
interpretable as functional or structural constraints and show that our simple
and unsupervised method has an accuracy similar to state-of-the-art methods,
including supervised ones. 
\end{onecolabstract}
\section{Introduction}

As of January 2017, the UniProt database \citep{magrane2011uniprot} lists 75.431 amino
acid variants of their sequences and classifies 28.891 of them as
associated with human diseases, many of which lethal. Since every
human genome is estimated to contain 74 de novo SNVs and around 0.6 de
novo 
missense mutations
\citep{veltman2012novo}, the number of known variants and the need for their
classification is likely to increase strongly in the near future due to cheap
sequencing and the trend towards \textit{personal genomics}
\citep{angrist2016personal}. The potential of the field can be
appreciated considering private companies like 23andme, which
routinely publish genome-wide association studies based on hundreds of
thousands of human genomes
\citep{chen2016analysis,okbay2016genetic,okbay2016genome,strachan2016meta}.
Problems arise when considering rare variants or experiments with small sample
sizes, where the statistics remains insufficient to conclusively label a
specific variant or a gene as disease-associated \citep{purcell2014polygenic}.
Another problem is that the sheer number of variants discovered prohibits a
further experimental analysis of most of them. In both cases computational
methods can help: They can be used for extracting interesting variants from a
pool of unknown pathogenicity; and sometimes used to understand the etiology of
the disease with which the variant is associated \citep{wang2012two}.  Many
computational methods for predicting pathogenicity of amino acid substitutions
exist. Unsupervised methods look at different proxies for pathogenicity and are,
for example, based on the frequencies of amino acids at the mutated residue in a
multiple sequence alignment of homologous sequences \citep{kumar2009predicting},
or on the time of conservation of the mutated residue in a reconstructed
phylogenetic tree \citep{tang2016panther}. Supervised methods combine such
proxies and add other information like structural data or annotations
\citep{adzhubei2013predicting,kircher2014general}. Unsurprisingly, they often
perform better than unsupervised methods, but arguably not by a very large
margin \citep{tang2016panther,grimm2015evaluation}.  Inspired by the successes
of global probability models for protein sequences in fields like the
prediction of protein residue contacts
\citep{weigt2009identification,Jones,morcos2011direct,Kamisetty2013}, 
the inference of protein interaction
networks \citep{ovchinnikov2014robust,hopf2014sequence,feinauer2016inter} and
the modeling of mutational landscapes in bacteria and viruses
\citep{chakraborty2014hiv,morcos2014coevolutionary,figliuzzi2015coevolutionary},
we propose in this work the inclusion of a
new type of information when predicting the pathogenicity of mutations in
humans: the sequence context in which the mutation appears. We first show that
the sequence context indeed carries information useful in the prediction of
pathogenicity by comparing the performance of two sequence-based models, one of
which includes the context dependence and the other one not. We show cases in
which this context dependence is interpretable as structural or functional
constraints that the model exploits to correctly predict the pathogenicity of
the mutation. In the age of machine learning, where results of algorithmss are often
hard to interpret, this might be an important advantage of a method. We then
go on and show that the performance of the method compares favorably with
published methods, supervised and unsupervised.

\section{Results}

\vspace{0.5cm} 
\textbf{Outline}
We assess the pathogenicity of a mutation by first mapping the mutated residue
to a consensus position in one of the profile hidden Markov models (pHMM) \citep{durbin1998biological}
provided by the protein domain family database (Pfam) \citep{finn2016pfam}. We then fit a maximum
entropy probability distribution to the multiple sequence alignment (MSA) of
this protein domain family and calculate scores for pathogenicity based on this
probability distribution.  In order to assess whether the sequence context in
which a mutation occurs carries information about its pathogenicity, we compare
the predictive power of scores based on a context-dependent model with scores
based on a context-independent model (short \textit{independent model}). The
scores reflect this context dependence or context-independence: Changing the
sequence context in which the mutation occurs changes the score derived from the
context-dependent model, while the score derived from the context-independent
model remains the same.
\vspace{0.5cm} 

\textbf{Independent Model}
The independent model we use is a maximum-entropy distribution that reproduces
amino acid frequencies in the MSA while ignoring covariances between amino
acids. This model treats all residues in the MSA as independent and thus
neglects any possible epistatic effects (see \textit{Materials and Methods}).
The probability distribution is the same as in profile models used for
constructing MSAs, and for aligned positions in profile
hidden Markov models when neglecting gaps
\citep{durbin1998biological}. Although such models cannot capture 
structural or functional constraints comprising more than one residue, they have
a major advantage in terms of simplicity: The number of parameters scales
linearly with the length of the sequences (see \textit{Materials and Methods}),
which means that relatively few sequences are needed to fit the model well.

\vspace{0.5cm} 
\textbf{Context-Dependent Model}
Context dependence can be included in our framework by using a probability
distribution in which the residues in a protein sequence are not independent
from each other. This means that the probability of finding an amino acid at
some specific position depends on the amino acids found at other positions.
Scores that quantify the pathogenicity of mutations based on such probability
distributions show epistatic effects and the predicted pathogenicity of a
mutation depends on the sequence context. In this work we use the
direct-coupling analysis (DCA) \citep{weigt2009identification}, which constructs a 
maximum-entropy model constrained to reproduce the covariances between amino
acids in the MSA. It includes context dependence and has been used successfully
for modeling protein sequences (see \textit{Materials and Methods} and
\citealp{stein2015inferring} for a general introduction). The inference of the
model follows \citealp{ekeberg2013improved}.  The context-dependent model
should be able to capture structural or functional constraints comprising more
than one residue, but its parameters scale quadratically with the length of the
protein (see \textit{Materials and Methods}). We show below that the available
sequences are nonetheless sufficient to make the context-dependent model
perform better than the independent model. A feature of the context-dependent
model is that the influence of the sequence context is easily analyzed: Part of
the change in log probability when substituting an amino acid in a sequence is
made up of contributions from other residues. We call these contributions $c_j$, where $j$ labels a non-mutated residue.
We show below that the $c_j$ can be used to interpret the score (see \textit{Materials and Methods}).

\vspace{0.5cm} 
\textbf{Scoring Pathogenicity} A natural score for the pathogenicity of a
mutation is the difference in the logarithm of the probabilities of the mutated
and the original sequence \citep{figliuzzi2015coevolutionary}. We call this
score $\Delta L$.  We show below that this quantity gives good results, but is
not optimal. This is probably due a dependence of the expected value of the
difference in the logarithms on the length of the protein domain and the
sampling depth of the original MSA. This makes a comparison of this score
across different protein domain families problematic.  We therefore propose
another measure \textit{r} that is based on a imaginary mutagenesis experiment:
We calculate $\Delta L$ for all possible single-site amino acid substitutions
in the original sequence and determine which rank the mutated sequence to be
assessed has in this spectrum. We then map this rank to the interval between 0
and 1 by dividing by the number of different $\Delta L$ in the hypothetical
experiment.  We show below that this score outperforms $\Delta L$
significantly.  We did not attempt to define thresholds for classification of
mutations as deleterious or benign.  Such a classification should be addressed
in a supervised framework using $r$ as an additional input. This we reserve for
future investigations. Throughout the rest of the paper, we call $r$ and $\Delta L$ the scores calculated on the context-dependent model, and $r^{IM}$ and $\Delta L^{IM}$ the scores calculated on the independent model.

\begin{figure}
\centering
\includegraphics[width=\textwidth]{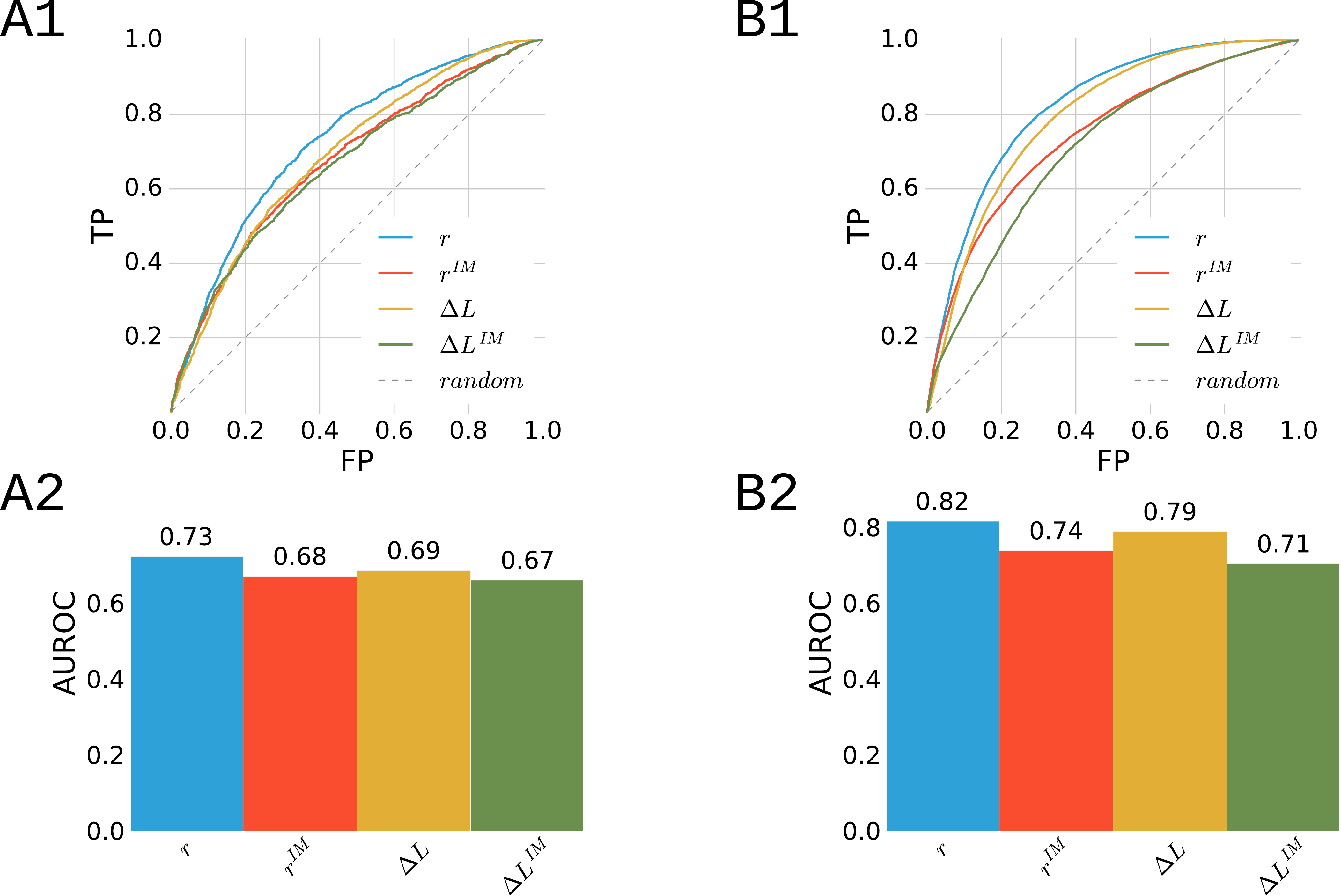}
\caption{\textbf{A1,A2:} ROC curves (A1) and area under the curves (A2) for scores from
the context-dependent ($\Delta L$ and $r$) and the independent model ($\Delta
L^{IM}$ and $r^{IM}$), calculated on the \textit{SwissVarSelected}
\citep{grimm2015evaluation} dataset where the method returned predictions. TP
is true positive rate and FP false positive rate. \textbf{B1,B2:} ROC curves
(B1) and area under the curves (B2) for scores from the context-dependent
($\Delta L$ and $r$) and the independent model ($\Delta L^{IM}$ and $r^{IM}$),
calculated on all mutations annotated in UniProt \citep{magrane2011uniprot}
where the method returned predictions. TP is true positive rate and FP false
positive rate.}
\label{fig:context_vs_ind}
\end{figure}

\vspace{0.5cm}
\textbf{Comparison between Context-Dependent Model and Independent Model} We
calculated the scores $r$,$r^{IM}$,$\Delta L$ and $\Delta^{IM}$ for the
\textit{SwissVarSelected} \citep{grimm2015evaluation} benchmark set. Of the
11028 mutations in this dataset that we could map to a UniProt accession number,
about 49\% could be mapped to a consensus position in a Pfam domain family. We
calculated the receiver operating characteristic curve
\citep{friedman2001elements} for the various scores and show them them in Fig.
\ref{fig:context_vs_ind}. We first notice that the scores based on imaginary
mutagenesis experiments, $r$ and $r^{IM}$ perform better than their
counterparts $\Delta L$ and $\Delta L^{IM}$. We furthermore notice that the
scores derived from context-dependent models, $r$ and $\Delta L$ outperform the
scores derived from independent models, $r^{IM}$ and $\Delta L^{IM}$. The
best-performing score is $r$. We therefore conclude that context dependence
represents indeed additional and useful information for the prediction of
pathogenicity. To further corroborate these results we repeated the test on all
annotated mutations and polymorphisms in the Uniprot Database
\citep{magrane2011uniprot}. While the overall performance of all models  on
this dataset is higher than on the \textit{SwissVarSelected} dataset (in line
with \citealp{tang2016panther}), the relative performances remain approximately
equal (see Fig. \ref{fig:context_vs_ind}).

\begin{figure}
\centering
\includegraphics[width=\textwidth]{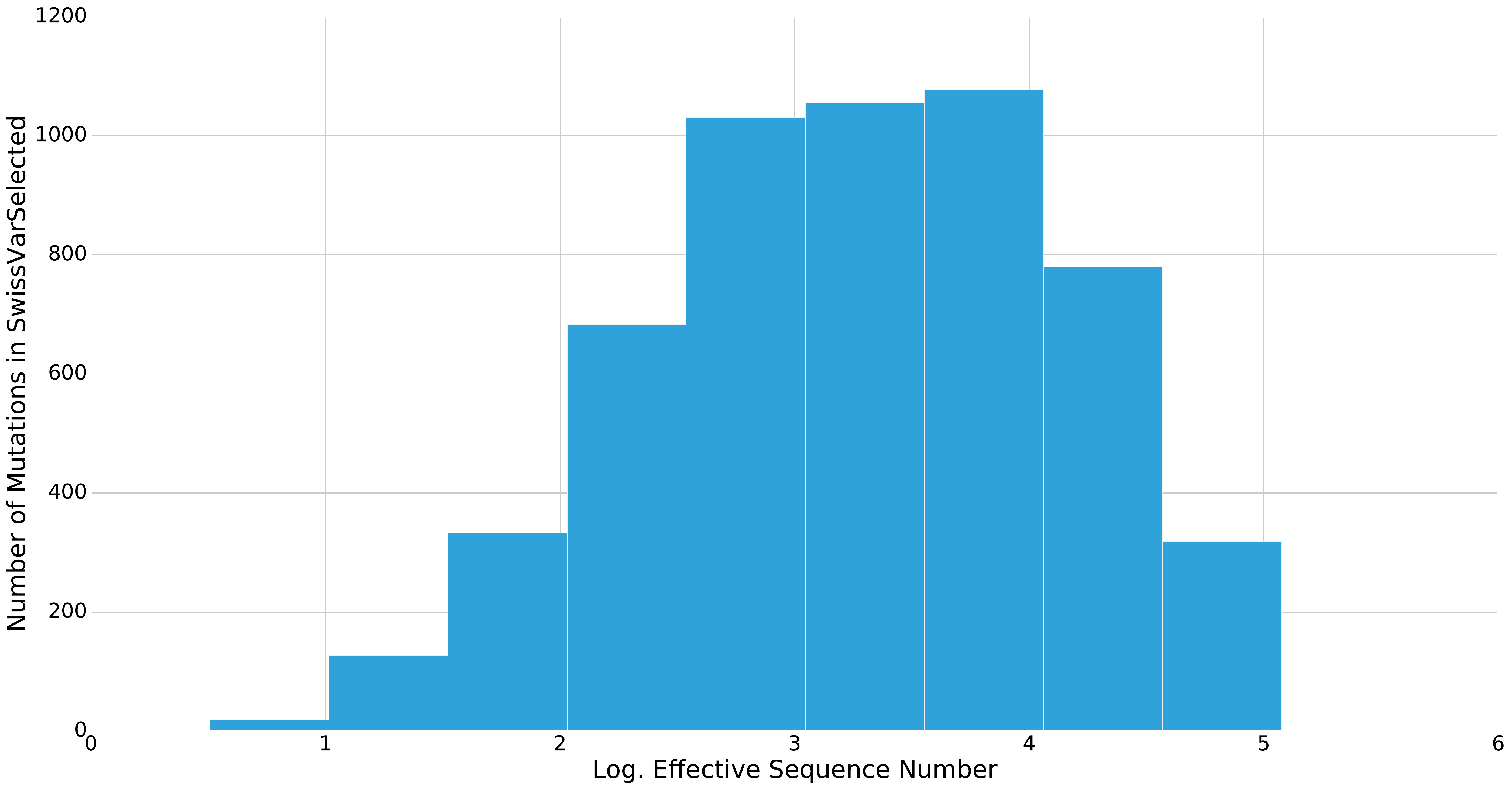}
\caption{Histogram of the decadic logarithm of the effective sequence numbers in the multiple sequence alignments corresponding to the mutations in the \textit{SwissVarSelected} dataset \citep{grimm2015evaluation}. Only mutations that could be mapped to a Pfam family are included.}
\label{fig:meff_histo}
\end{figure}

\begin{figure}
\centering
\includegraphics[width=\textwidth]{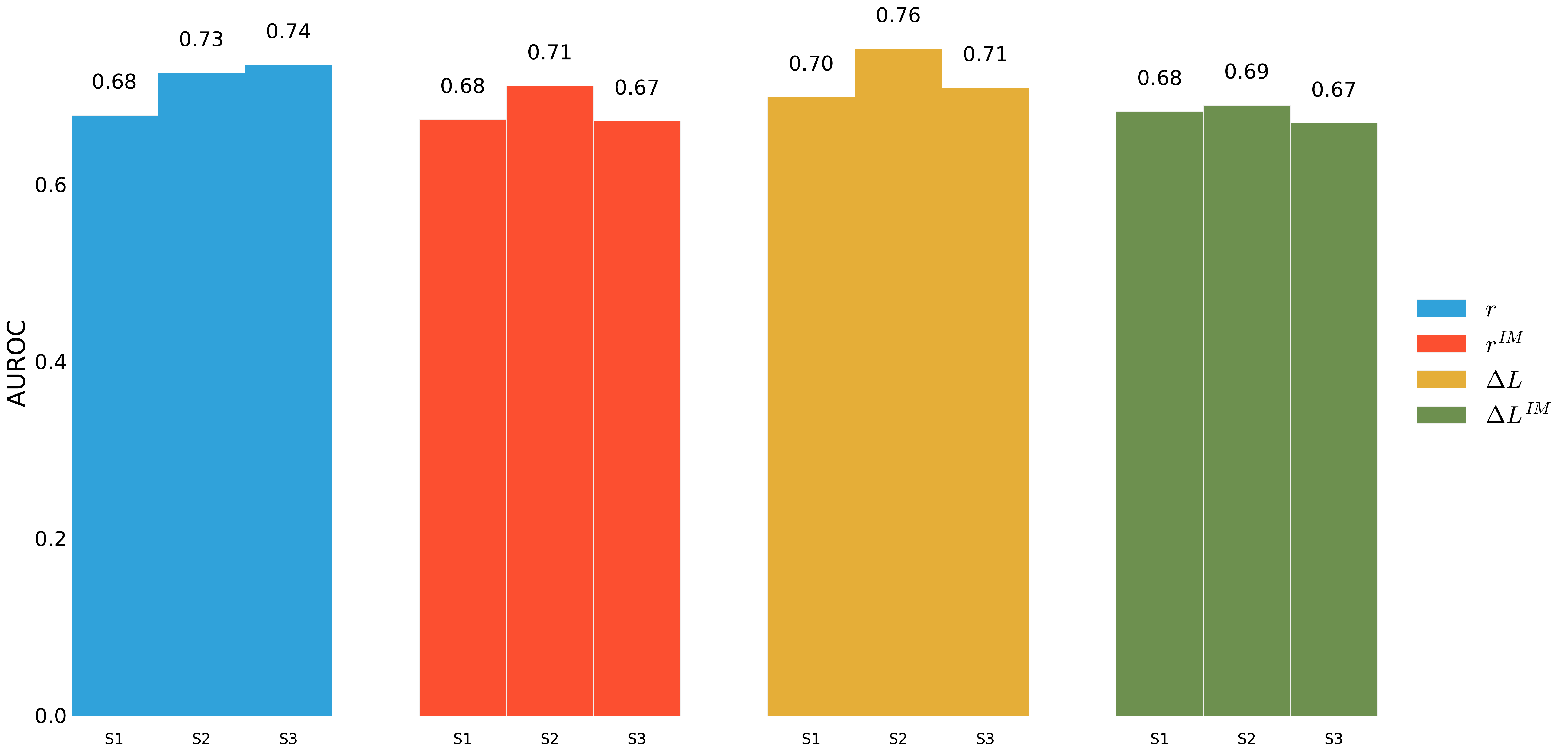}
\caption{Bounds are: S1 0 - 139, S2 139 -  346, S3 346 - 749}
\caption{Performance (AUROC) of $r$ score on subsets of
\textit{SwissVarSelected}. Subsets are labeled S (mutations with effective
number of sequences less than 139), M (between 139 and 346 effective
sequences) and L (between 346 and 749 sequences). To evaluate performance
the subsets were downsampled to arrive at the same fraction of
pathogenic/non-pathogenic mutations (40\% to 60\%) as in \textit{SwissVarSelected}.}
\label{fig:meff_lower_terciles}.
\end{figure}

\vspace{0.5cm}
\textbf{Sampling Depth} 
Since the context-dependent model is based on a large number of parameters (see
\textit{Materials and Methods}), it is important to have a sufficient number of
homologous sequences in the model inference process. We used the
\textit{effective number of sequences} in the MSA as calculated in the
inference process (see \citealp{ekeberg2013improved}) to estimate the amount of
information in the MSA. This differs from the number of sequences in the
alignment because of corrections for phylogenetic and experimental biases
\citep{ekeberg2013improved}. The mean number of effective sequences for the
mutations in the \textit{SwissVarSelected} that could be mapped to Pfam
families is 9018 sequences, with a median of 1923 (a histogram of the number of
effective sequences can be found in Fig. \ref{fig:meff_histo}). While this
average would be considered as sufficient for structural predictions
\citep{morcos2011direct}, we also observe a large number of mutations that
correspond to multiple sequence alignments with less than 500 effective
sequences. This would be considered as not sufficient in structural
predictions. Since is not clear how the performance in structural prediction
relates to the performance in the prediction of pathogenicity, we quantified the
effect of sampling depth on predictive performance.  We divided the
\textit{SwissVarSelected} dataset in three approximately equal parts by the
number of effective sequences in the corresponding Pfam families. The
corresponding thresholds were 754 and 5828 effective sequences.  We then
calculated the AUROCs for the context-dependent model and the independent model
after sampling from each subset to reach the same proportion of pathogenic and
non-pathogenic variants as in the original dataset (approximately 40\% to
60\%).  To our surprise, the difference in performance when using $r$ as a
score was not very pronounced in the three subsets (see Fig. \ref{fig:meff} in
the \textit{Supplemental Material}).  We take this as evidence that even for
the tercile with the least sequences the number of sequences is sufficient. We
also observed that for the largest tercile the performance worsens and for the
independent model the performance seems to generally deteriorate with higher
sequences numbers (see \ref{fig:meff} in the \textit{Supplemental Material}).
This could be either due to systematic problems with the largest families, like
false labeling or non-conservative inclusion thresholds of the MSAs, or to
random variability in the data set.  A further partitioning of the smallest
tercile shows that the performance with between 139 and 346 sequences is only
marginally worse than with more than 346 sequences  (see Fig.
\ref{fig:meff_lower_terciles} the \textit{Supplemental Material}). A minimum of
139 sequences includes 89\% of the mutations in the test set (see Fig.
\ref{fig:meff_histo}).  The general outlook seems therefore to be that for most
mutations that can be mapped to a position in a PFAM domain the number of
sequences is sufficient. The problem that many mutations cannot be mapped to a
Pfam domain (about 51\% in the \textit{SwissVarSelected} dataset) is likely to
pose a larger challenge in actual usage than lacking sampling depth.

\begin{figure}
\includegraphics[width=\textwidth]{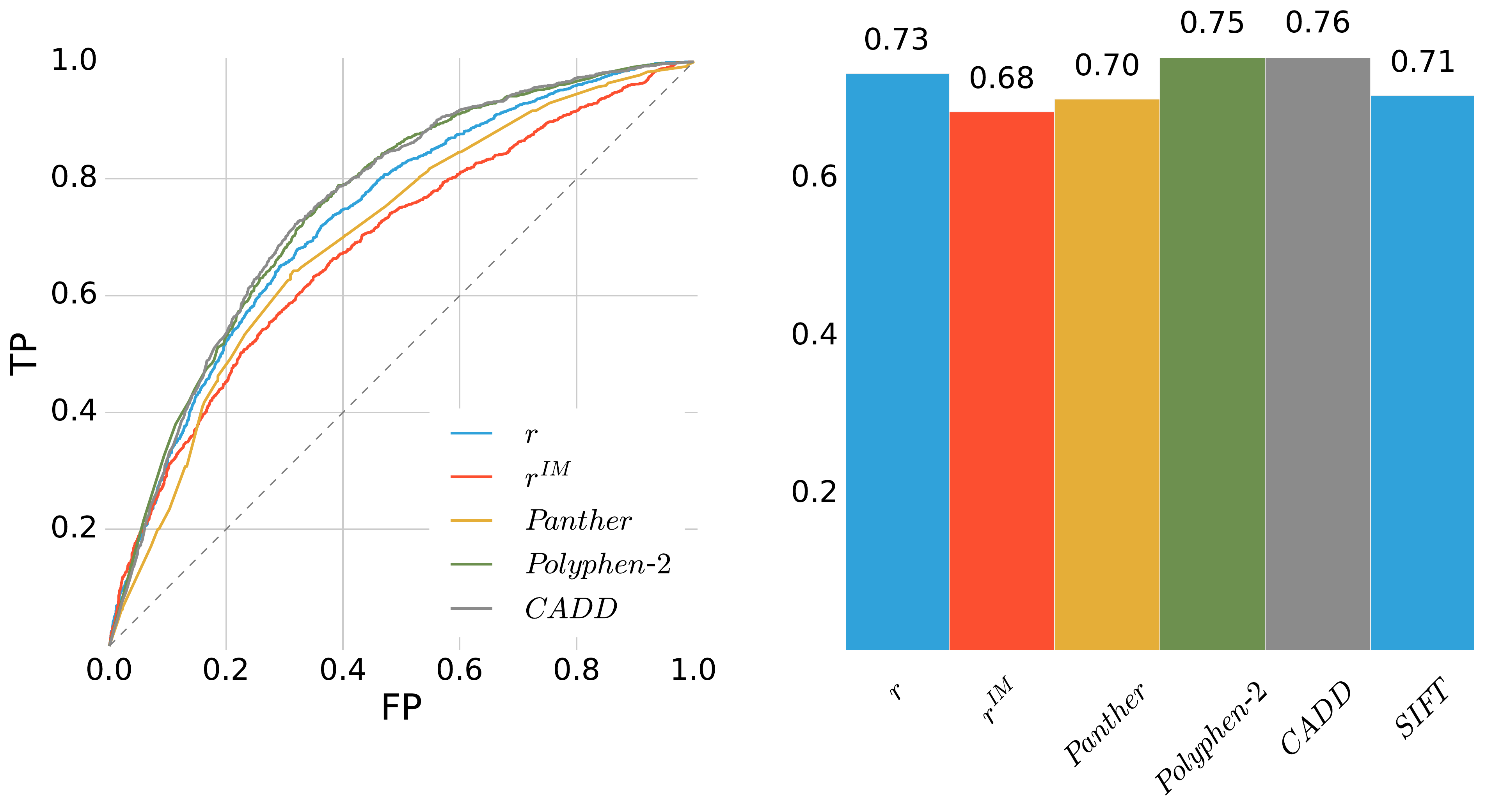}
\caption{ROC curves (left) and area under the curves (right) for scores from
the context-dependent model ($r$) and the independent model ($r^{IM}$) in
comparison with other methods as labeled. The results for the other methods
were obtained as described in \textit{Materials in Methods}. The dataset used
here is \textit{SwissVarSelected} \citep{grimm2015evaluation} and only
mutations where kept where all methods returned predictions. TP is true positive rate and FP is false positive rate.}
\label{fig:context_vs_other}
\end{figure}

\vspace{0.5cm} 
\textbf{Comparison with other Methods} The central message of this work is that
context dependence is important for predicting the pathogenicity of amino acid
substitutions. It is nonetheless interesting to compare the performance of our
method to state-of-the-art predictors. As one would expect the best performing
methods today are supervised methods that take information from different
sources like sequence data or experimental structures into account
\citep{grimm2015evaluation}. In Fig.  \ref{fig:context_vs_other} we compare our
context-dependent model and the independent model to: Panther-PSEP, a recent
unsupervised method based on phylogeny \citep{tang2016panther}; SIFT, a highly
cited unsupervised method using residue conservation as a measure for
pathogenicity \citep{kumar2009predicting}; Polyphen-2, a supervised method
using 11 features derived from homologous sequences and structural data
\citep{adzhubei2013predicting}; CADD, a supervised method for genetic variants
using 63 annotations as inputs, including protein level annotations derived
from other pathogenicity-predictors like Polyphen or SIFT
\citep{kircher2014general}. In Fig. \ref{fig:context_vs_other} we show the
AUROC values achieved by the different methods. All methods have AUROCs around
0.7, with supervised methods slightly outperforming the unsupervised methods.
It is surprising that the large number of features (including structural
features) that the supervised methods exploit do not lead to a larger boost in
performance, but this has been observed before \citep{thusberg2011performance}.
While the unsupervised methods generally show a very similar performance, we
observe by comparing the performance of $r$ and $r^{IM}$ that the inclusion of
the context dependence lifts the performance of our model from the worst-performing
unsupervised method to the best-performing unsupervised method presented.  As a
last point we notice that the performance of PANTHER in our runs is at odds
with the performance reported in \citealp{tang2016panther} (area under the
curve 0.72 vs. 0.70). We speculate this to be a data artifact: Since we had
several methods in the comparison, we reduced the \textit{SwissVarSelected}
dataset to only the mutations for which all methods returned results. This left
only 3857 of 12739 mutations; and since the authors in
\citealp{tang2016panther} presumably did the same, the overlap between the two
datasets used might be even smaller. 

\section{Discussion}

In this work, we have shown that including context dependence in methods for
predicting the pathogenicity of amino acid substitutions can lead to a
significant boost in performance. We have also shown that a relatively simple
unsupervised method based on maximum entropy modeling and using no biological
prior information can compete with state-of-the-art unsupervised and supervised
methods. 

One of the characteristics of the probabilistic model
constructed by DCA is the explicit estimation of epistatic 
couplings between different residues. This allows the disentangling of
the context dependence into contributions from individual residues in
the unmutated sequence, and to \textbf{interpret
  context dependence} in a number of sample cases. In this respect, we
speculate that the 
context dependence helps predicting the pathogenicity of mutations because the
corresponding probability distribution captures constraints acting on groups of
residues, i.e.  residues that are part of a functional group or form structural
bonds with each other. In order to validate this point, we took a closer look
at mutations that were labeled as disease-associated in the dataset, had a high
$r$-score in the context-dependent model and a low $r^{IM}$ score in the
independent model. We found 40 pathogenic mutations
for which $r-r^{IM}>0.5$. We then checked existing literature on these
mutations for examples in which our models could be interpreted consistently
with this literature.
We found evidence for at least three cases in which context dependence can lead
to better performance.  The three cases are 1) mutations within functional
groups, 2) mutations leading to a loss of a structural contact, 3) mutations
that are common in one cluster of sequences appearing in another cluster.  We
elaborate these cases with examples. 

\begin{figure*}
\centering
\includegraphics[width=1.0\textwidth]{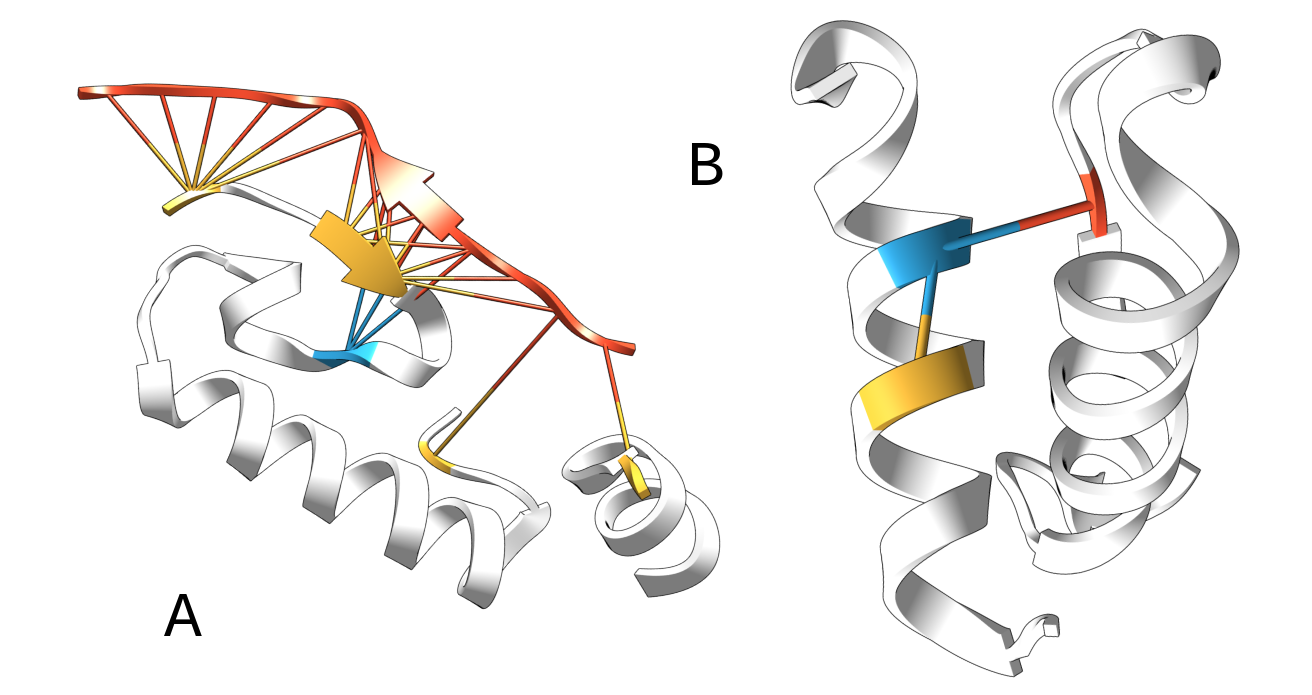}
\caption{\textbf{A} Interaction between PRKACA (lower left) and its regulator
PRKAR1A (upper right, red). The yellow residues are the strongest contributors
to the drop in $\Delta L$ when exchanging an arginine with a leucine at residue
206 of PRKACA (blue). The sticks represent residue contacts between PRKACA and
its regulator. The PDB used is 3TNP \citep{zhang2012structure}. This figure
shows only fragments for clarity. See \textit{Supplemental Material} for the full
structure. \textbf{B} Interaction between two $E1$-$\beta$-subunits within the
BKCD complex. Sticks represent the two strongest contacts by $F^{APC}$ score of
residue 346 (blue).  Only residue 346 of the left subunit and its contacts are
colored: an $\alpha$-helix intra-chain contact with residue 342 (yellow) and a
homomeric contact with residue 357 of the right chain (red). A trivial contact
with residue 347 is not shown. The PDB used is 1X7Y \citep{wynn2004molecular}.
This figure shows only fragments for clarity. See \textit{Supplemental
Material} for the full structure.}
\label{fig:cushing_maple}
\end{figure*}

\vspace{0.5cm}
\textbf{Adrenal Cushing's Syndrome} Patients with adrenal Cushing's syndrome
exhibit a variety of symptoms stemming from an overexposure to cortisol
\citep{goh2014recurrent}. A specific mutation that has been found in such
tumors is PRKACA\textsuperscript{L206R}, affecting a subunit of a protein
kinase. This mutation prevents that PRKACA is bound by its regulator PRKAR1A,
increasing its phosphorylation activity and leading to cortisol production
downstream \citep{goh2014recurrent}.  Since both arginine and leucine are
common at this position in the corresponding protein domain family, the
independent model assigns both sequences similar rank ($r^{IM} = 0.14$ for
PRKACA and $r^{IM} = 0.153$ for PRKACA\textsuperscript{L206R}.  In contrast,
the context-dependent model scores the mutated sequence considerably worse than
the original sequence ($r=0.02$ for PRKACA and $r=0.73$ for
PRKACA\textsuperscript{L206R}). In order to understand this, we calculated the
contributions from other residues to the corresponding drop in $\Delta L$ in
the context-dependent model when exchanging the arginine with the leucine (see
\textit{Materials and Methods}). This contribution quantifies how strongly the
specific amino acid found in some other position favors or disfavors the
mutation. In Fig. \ref{fig:cushing_maple}A we color residue 206 in blue, the 4
strongest contributors in yellow and the regulator of PRKACA in red.  Of these
4 only two, 201G and 200C, qualify as contacts with 3.7 {\AA} and 3.2 {\AA}
minimal heavy atom distance to 206L (distances are extracted from the
PDB 3TNP \citep{zhang2012structure} of a mouse homolog of the human
protein).  The other two, 197W and 238P, are more distant on the chain
and also more distant in the structure with 10 {\AA} and 8.9 {\AA}
minimal heavy atom distance to 
206L.  However, all four are in the interaction surface between PRKACA and an
11-residue segment (95R to 106T) of its regulator PRKAR1A, with 21 contacts
between this segment and the 4 contributors and 206L.  We take this as evidence
that the context-dependent model has captured a multi-residue constraint acting
on residues within the functional group responsible for the interaction. This
functional group represents the sequence context that allows the correct
classification of this mutations as pathogenic. This is in full agreement with
the idea that the mutation impairs the binding of the regulator PRKAR1A to
PRKACA \citep{goh2014recurrent}.

\vspace{0.5cm} 
\textbf{Classic Maple Syrup Urine Disease} The disease is caused by an impaired
metabolism of branched-chain amino acids \citep{chace1995rapid}.  Carriers
often die young, within months or days after birth \citep{wang2012two}.  The
mutation BCKDHB\textsuperscript{Q346R} has been implicated in the disease
\citep{wang2012two}. The BCKDHB gene encodes the two $\beta$-subunits of the
$E1$-complex, which is part of the multi-subunit enzyme complex BCKD involved
in the metabolism of leucine, isoleucine and valine. \citep{indo1987altered}.
There is evidence that the mutation destabilizes the interaction of 346R with
357I between the $\beta$-subunits \citep{wang2012two}. The two residues are
2.72 {\AA} apart in the complex (minimum distance between heavy atoms, PDB 1X7Y
\citep{wynn2004molecular}).  The rank of the mutated sequence is considerably
worse in the context-dependent model ($r = 0.83$) than in the independent model
($r^{IM} = 0.27$). We speculated that this might be due to a structural
constraint between residue 346 and residue 357 that the independent model
cannot capture. Shifting to structural analysis, we calculated $F^{APC}$ scores
for the residues (see \textit{Materials and Methods}). These are
quantities determined by the model that can be used to infer residue contacts
\citep{ekeberg2013improved}. We found that the residue with the strongest coupling
to residue 346 was indeed residue 357 (excluding short-range couplings),
indicating a strong interaction between the residues in the context-dependent
model (see Fig.\ref{fig:cushing_maple}B). We therefore conclude that the
context-dependent model captured a structural constraint involved in the
homo-dimerization of the two $\beta$-subunits. The sequence context that helped
the context-dependent model to asses pathogenicity in this interpretation is
reduced to a single residue.

\vspace{0.5cm} 
\textbf{Alpha-Mannosidosis} The disease is caused by a deficient catabolism of
N-linked oligosaccharides, due to a defective enzyme lysosomal
alpha-mannosidase encoded in the gene MAN2B1. Patients might suffer from
symptoms like mental retardation, changed facial features and impaired hearing
abilities \citep{roces2004efficacy}. A mutation implicated in the disease is
MAN2B1\textsuperscript{R202P} \citep{stensland2012identification}. It has a
high pathogenicity-score in the context-dependent model ($r = 0.89$), but a low
score in the independent model ($r^{IM} = 0.14$).  Calculating the largest
contributors (see \textit{Materials and Methods}) to the change in the score,
we find contributions of similar magnitude from many residues. An analysis of
PDB 1O7D \citep{heikinheimo2003structure} revealed that only 4 of the ten
largest contributors are in contact with residue 202. Furthermore, these 4 are
all less than 8 residues apart from it. A possible explanation for such cases
is that the sequences cluster and the mutation exchanges an amino acid common
at the residue in one cluster with an amino acid common in another cluster. The
context-dependent model might not capture biologically significant constraints
in such cases, but can nonetheless single out residues that are uncommon in the
specific context.  In fact, we observe that sequences that have an arginine at
residue 202 are significantly more similar to each other (mean Hamming distance
0.51) than to sequences that have a proline at that residue (mean Hamming
distance 0.74). This can be summed up by saying that the sequence context that
helped the context-dependent model to assess the pathogenicity in this case is
the entire protein sequence.  This is of course a hypothesis of last resort.
Another plausible explanation is that there is biological constraint captured
by the context-dependent model, but we are not able to discover it.

\vspace{0.5cm}

{\bf Note:} While finalizing the redaction of this article, another article
with partially overlapping results \citep{hopf2017mutation} has been published
online. The paper comes to a similar conclusion, i.e. the inclusion of
sequence context dependencies improves the prediction of mutation effects. While
a direct comparison is difficult since the paper uses different data sets, it
is built upon the $\Delta L$ log-probability score, which on all our data-sets
performs significantly worse than the $r$ rank score proposed in our work. We
also note that our main data-set, \textit{SwissVarSelected}, was explicitly
designed as a comparable benchmark and used as such in recent publications
\citep{grimm2015evaluation,tang2016panther}. We therefore believe that our work
gives additional and valuable insights on how modeling context dependence
affects predictive performance compared to established methods.

\section{Materials and Methods}

\vspace{0.5cm}
\textbf{Outline of the Method} A mutation is assumed to be given as a Uniprot
Accession Number, a position, a reference amino acid and a substituted amino
acid. The Pfam domain architecture of the corresponding Uniprot sequence is
extracted from the Pfam database \citep{finn2016pfam} and it is checked
whether the mutated residue can be mapped to a consensus column of a Pfam pHMM.
If not, the method does not return a result. If yes, plmDCA is run on the Pfam
MSA of the protein domain family and the various scores are calculated with its
output. For the independent model the inference with plmDCA is replaced by a
analytic formula.

\vspace{0.5cm}
\textbf{Context-Dependent Model}
The context-dependent model defines a maximum entropy probability distribution
$p(a)$ for all amino acid sequences of length $N$ \citep{morcos2011direct}. The
model can be derived by enforcing the constraint that it should reproduce the
covariances between amino acids in the MSA \citep{stein2015inferring}. The
logarithm of this distribution can be written as

\begin{equation}
	\log p(a) = \sum\limits_{i<j} J_{ij}(a_i,a_j) + \sum\limits_{i} h_i(a_i) - \log Z.
\end{equation}

Here, $a_i$ is the amino acid at residue $i$, the fields $h_i(a)$ represent the
propensity of position $i$ to be amino acid $a$ and $J_{ij}(a,b)$ are coupling
parameters that represent the context dependence in the model. $\log Z$ is a
normalization constant.
The parameters are inferred using plmDCA as described in \citep{ekeberg2013improved}.

An intuitive score for the pathogenicity of substituting amino acid $a_i$ with
$a_i'$ in sequence $a$ is the change in log probability \citep{figliuzzi2015coevolutionary}

\begin{equation}
	\Delta L  = h_i(a_i) - h_i(a_i') + \sum\limits_{j} c_j.
\end{equation}

The $c_j$ measure the contribution of residue $j$ to the score,
\begin{equation}
c_j = J_{ij}(a_i,a_j) - J_{ij}(a_i',a_j). 
\end{equation}
However, the model for a protein domain of even medium length $N=50$ has
already more than $5 \times 10^5$ coupling parameters, which are often inferred
on MSAs of 1000 sequences or less. We expect this measure therefore to be very
noisy. 

A general measure for the interaction between $i$ and $j$ are the $F^{APC}$
scores used in structure prediction, which can be calculated from the $J_{ij}$
\citep{ekeberg2013improved}.   They contain a sum over many couplings and we
expect them to be more stable than the $c_j$. Since the $F^{APC}$ scores have been
shown to capture structural contacts, we expect them to be useful when the
substitution of an amino acid violates as structural constraint.

Another measure for the pathogenicity of a mutation is the rank of the mutated
amino acid sequence in a hypothetical mutagenesis experiment. To this end, we
calculated $\Delta L$ for all possible single amino acid substitutions in the
original sequence. We then divided the rank of the mutated sequence in this list
with the number of unique $\Delta L$ values in the hypothetical experiment. The
resulting score $r$ lies between 0 and 1 and is independent of the scale of
$\Delta L$. 

\vspace{0.5cm}
\textbf{Independent Model}
The independent model defines another maximum entropy probability distribution.
It can be derived by enforcing the constraint that the model should reproduce
the amino acid frequencies in the MSA. It does not contain any context
dependence. The log probability can be written as

\begin{equation}
\log p^{IM}(a) = \sum\limits_{i} \log p^{IM}_i(a_i) = \sum\limits_{i} h_i(a_i) - \log Z
\end{equation}

Using the same regularization (with regularization parameter $\lambda$ set to $0.01$) and reweighting techniques as in \citealp{ekeberg2013improved}, an analytical solution can be found to be

\begin{equation}
h_i(a) = \frac{f_i(a)}{2\lambda} - W\left( \frac{f_i(q)}{2\lambda} e^{\frac{f_i(a)}{2\lambda}} \right)
\end{equation}

where $f_i(a)$ is the reweighted frequency of amino acid $a$ at position $i$,
$q$ the most frequent amino acid at $i$ and $W$ the Lambert W function.
We define $\Delta L^{IM}$ and $r^{IM}$
analogously to above.

\vspace{0.5cm}
\textbf{Datasets and other Methods}
The dataset \textit{SwissVarSelected} has recently been shown to be a good
benchmark set for the prediction of pathogenicity \citep{grimm2015evaluation}.
We used it as the main dataset since it does not overlap with the training sets
of the supervised methods with which we compare our method. The dataset and the
results of the methods SIFT \citep{ng2003sift} and CADD
\citep{kircher2014general} have been taken from \citep{grimm2015evaluation}.
Polyphen-2 \citep{adzhubei2013predicting} has been run on the webserver
provided by the authors. Panther \citep{tang2016panther} has been run with the
standalone software provided by the authors. As an additional dataset we
downloaded all annotated Uniprot mutations from
\href{http://www.uniprot.org/docs/humsavar}{here}.
Pfam version 30.0 was used \citep{finn2016pfam}.

\section*{Acknowledgments}

We are grateful to Matteo Figliuzzi and Olivier Tenaillon for many
discussions. MW acknowledges funding by the ANR project COEVSTAT
(ANR-13-BS04- 0012-01).  

\section*{Additional information}

The author(s) declare no competing financial interests.

\bibliography{bibliography} 
\bibliographystyle{plainnat}

\FloatBarrier
\section{Supplemental Material}

\begin{figure}[!htb]
\includegraphics[width=\textwidth]{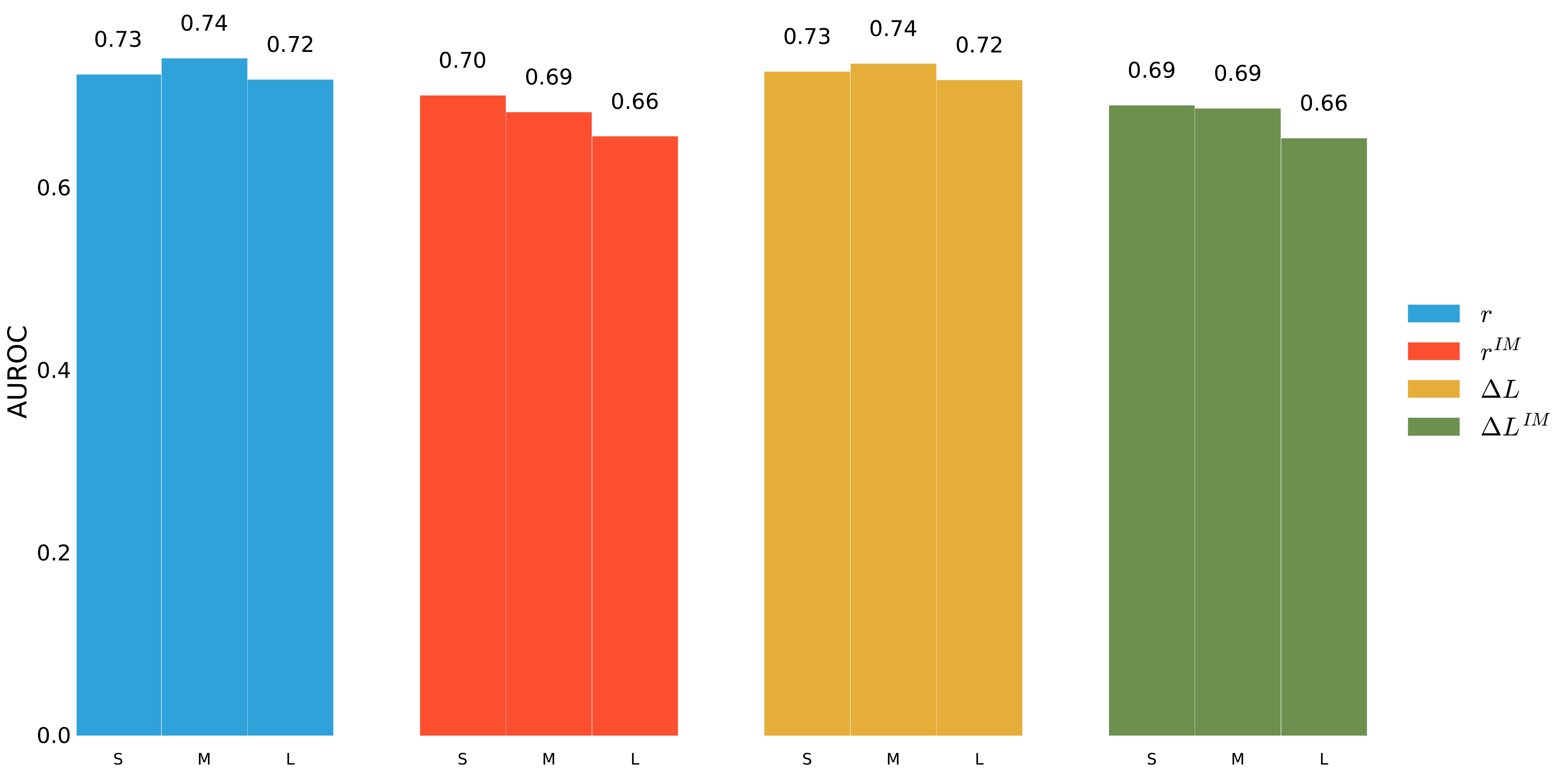}
\caption{Performance (AUROC) of $r$ score on subsets of
\textit{SwissVarSelected}. Subsets are labeled S (mutations with effective
number of sequences less than 754), M (between 754 and 5828 effective
sequences) and L (more than 5228 effective sequences). To evaluate performance
the subsets were downsampled to arrive at the same fraction of
pathogenic/non-pathogenic mutations (40\% to 60\%) as in
\textit{SwissVarSelected}.}
\label{fig:meff}
\end{figure}

\begin{figure}
\centering
\includegraphics[width=\textwidth]{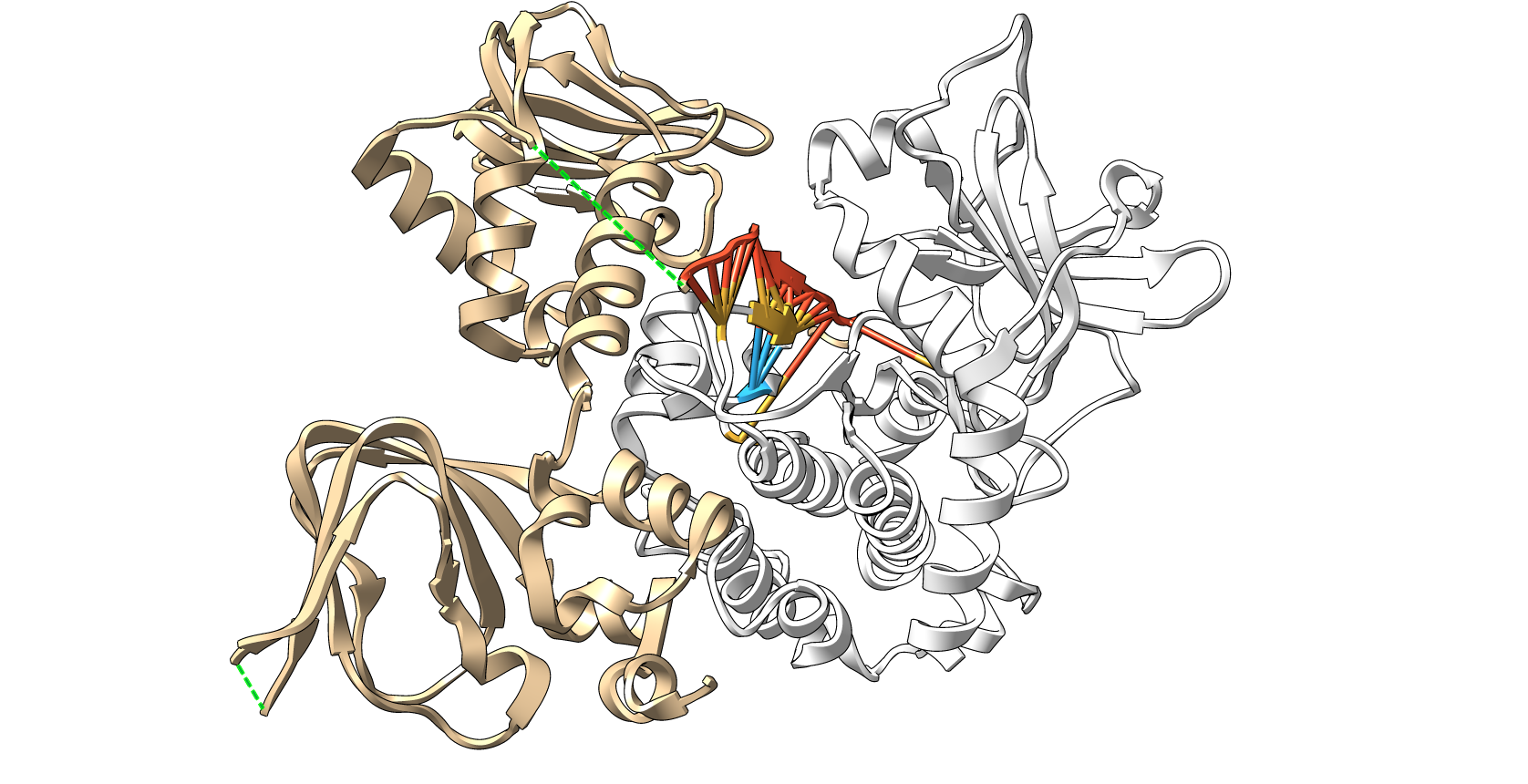}
\caption{Interaction between PRKACA (lower left) and its regulator
PRKAR1A (upper right, red). The yellow residues are the strongest contributors
to the drop in $\Delta L$ when exchanging an arginine with a leucine at residue
206 of PRKACA (blue). The sticks represent residue contacts between PRKACA and
its regulator. The PDB used is 3TNP \citep{zhang2012structure}.}
\label{fig:cushing_full}
\end{figure}

\begin{figure}
\centering
\includegraphics[width=\textwidth]{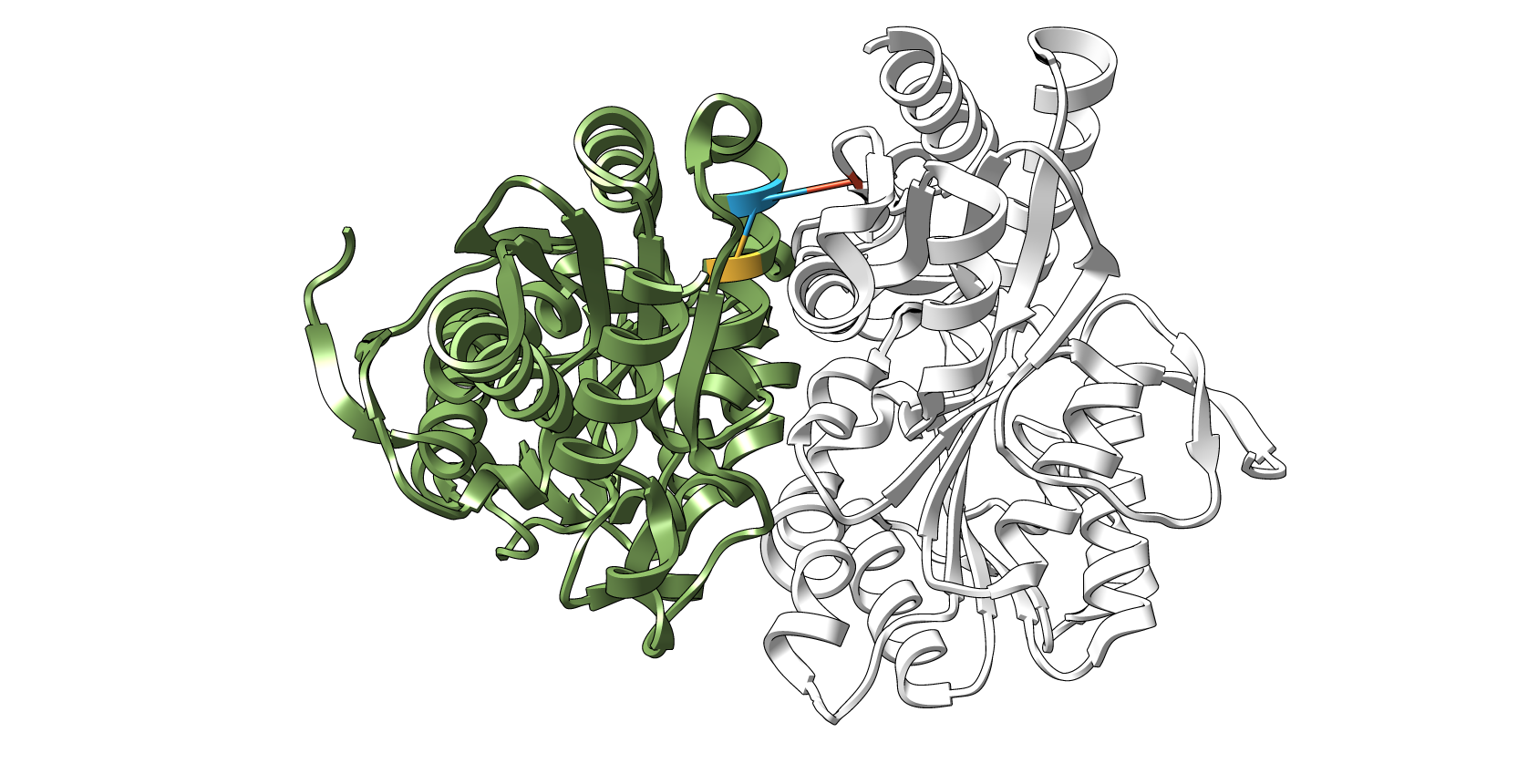}
\caption{Interaction between two $E1$-$\beta$-subunits (green and white) within the BKCD complex. Sticks
represent the two strongest contacts by $F^{APC}$ score of residue 346 (blue).
Only residue 346 of the left subunit and its contacts are colored: an
$\alpha$-helix intra-chain contact with residue 342 (yellow) and a homomeric
contact with residue 357 of the right chain (red). A trivial contact with
residue 347 is not shown. The PDB used is 1X7Y \citep{wynn2004molecular}.}
\label{fig:maple_full}
\end{figure}
\end{document}